\documentclass[12pt]{article}

\usepackage{amssymb,amsfonts,amsmath}
\usepackage{graphicx} 
\usepackage{indentfirst}
\usepackage{bbm}
\usepackage{color}
\usepackage{tikz}
\usetikzlibrary{snakes}
\usetikzlibrary{arrows}
\usetikzlibrary{calc}
\usetikzlibrary{patterns}

\parskip=\medskipamount

\newcommand {\cF}{{\cal F}}

\newcommand {\cH}{{\cal H}}

\newcommand {\cL}{{\cal L}}
\newcommand {\cM}{{\cal M}}
\newcommand {\cN}{{\cal N}}

\def\a{\alpha}

\def\d{\delta}

\def\G{\Gamma}

\def\q{\theta}

\def\z{\zeta}

\def\F{\Phi}
\def\J{\Psi}

\def\S{\Sigma}
\def\U{\Upsilon}

\newcommand{\ad}{{\dot{\alpha}}}                           %new
\newcommand{\ve}{\varepsilon}                            %new
\newcommand{\pa}{\partial}                           %new
\newcommand{\hf}{\frac12}
\newcommand{\sect}[1]{\setcounter{equation}{0}\section{#1}}

\newcommand{\be}{\begin{equation}}
\newcommand{\ee}{\end{equation}}
\newcommand{\bea}{\begin{eqnarray}}
\newcommand{\eea}{\end{eqnarray}}
\newcommand{\non}{\nonumber}

\def\dt#1{{\buildrel {\hbox{\LARGE .}} \over {#1}}}    % dot-over for sp/sb

\def\double #1{#1{\hbox{\kern-2pt $#1$}}}

\newcommand{\refer}[1]{(\ref{#1})}
\newcommand{\diff}{\mathrm{d}}

\newcommand{\Exp}[1]{\mathrm{e}^{#1}}

\newcommand{\oper}[1]{{\cal #1}}
\newcommand{\iunit}{\mathrm{i}}
\newcommand{\abs}[1]{\left|#1\right|}

\begin{document}

\begin{titlepage}

\begin{center}
{\Large \bf  Cotangent Bundle over}\\
{\vspace{3mm}}
{\Large \bf  Hermitian Symmetric Space $E_7/E_6 \times U(1)$ }\\
{\vspace{3mm}}
{\Large \bf  from Projective Superspace}
\end{center}

\begin{center}
{\large  
$^{a,b}$Masato Arai\footnote{masato.arai AT fukushima-nct.ac.jp}, 
$^{b,c}$Filip Blaschke\footnote{filip.blaschke AT fpf.slu.cz}
} \\
\vspace{5mm}

{\it 
$^a$Fukushima National College of Technology\\
Fukushima 970-8034, Japan\\
$^b$Institute of Experimental and Applied Physics, \\Czech Technical University in Prague\\ 
128 00 Prague 2, Czech Republic\\
$^c$Institute of Physics, Silesian University in Opava\\
746~01 Opava, Czech Republic\\
}

\vspace{2mm}

\end{center}
\vspace{5mm}

\begin{abstract}
\baselineskip=14pt
\noindent
We construct an $\cN=2$ supersymmetric sigma model on the cotangent bundle over the Hermitian symmetric 
space $E_7/(E_6\times U(1))$ in the projective superspace formalism, which is a manifest $\cN=2$ off-shell 
superfield formulation in four-dimensional spacetime. 
To obtain this model we elaborate on the results developed in arXiv:0811.0218 and present a new closed formula 
for the cotangent bundle action, which is valid for all Hermitian symmetric spaces.
We show that the structure of the cotangent bundle action is closely related to the analytic structure of 
the K\"ahler potential with respect to a uniform rescaling of coordinates.
\end{abstract}
\vspace{1cm}

\vfill
\end{titlepage}

\newpage
\setcounter{page}{1}
\renewcommand{\thefootnote}{\arabic{footnote}}
\setcounter{footnote}{0}

\tableofcontents{}
\vspace{1cm}
\bigskip\hrule

%\sect{Introduction}
%brabra

\sect{Introduction}
The number of supersymmetry is intimately related to the target space geometry in the supersymmetric sigma models 
\cite{Zumino}. The target spaces in the four-dimensional $\cN=2$ supersymmetric sigma models are hyperk\"ahler 
manifolds \cite{Alvarez-Gaume:1981hm}. Therefore supersymmetric field theory is a powerful tool to construct
new hyperk\"ahler manifolds. In order to construct new $\cN=2$ supersymmetric nonlinear sigma models, it is convenient 
to use a superfield formalism which makes supersymmetry manifest. Such an $\cN=2$ off-shell superfield formulation 
is the projective superspace formalism\footnote{Another $\cN=2$ off-shell superfield formulation
exists, which is called the harmonic superspace formalism \cite{GaIvKaOgSo}.} \cite{Karlhede:1984vr}. 
The projective superspace extends superspace at each point by an additional bosonic coordinate $\zeta$ which is 
a projective coordinate on ${\bf C}P^1$; action is written using contour integral over $\zeta$, and reality conditions 
are imposed, using complex conjugation of $\zeta$ composed with the antipodal map \cite{Karlhede:1984vr, LR1,LR2,G-RLRWvU}. 

In this formalism, there have been developments in construction of $\cN=2$ supersymmetric sigma models on the 
tangent bundles of the large class of the Hermitian symmetric spaces as well as using the generalized Legendre transform \cite{LR1}, 
the cotangent bundles corresponding to the hyperk\"ahler metrics \cite{GK1, GK2, ArNi, AKL, AKL2, KuNo}. The method 
used in \cite{GK1, GK2, ArNi, AKL} essentially rested on finding solutions to the $\cN= 2$ projective superspace auxiliary field
equations in K\"ahler normal coordinates at a point in the Hermitian symmetric space $\cM$ and then extending 
the solutions, using cleverly chosen coset representatives. This method was originally introduced in \cite{GK1, GK2} 
and was illustrated on the example of $\cM={\bf C}P^1$. The case of $\cM={\bf C}P^n$ was worked out in \cite{GK2, ArNi}. 
The classical Hermitian symmetric spaces as well as their non-compact versions were worked out in \cite{AKL}. The 
tangent bundle over $Q^n=SO(n+2)/(SO(n)\times U(1))$ was obtained in \cite{ArNi}.

Although the method in \cite{GK1, GK2, ArNi, AKL} is viable, it has a disadvantage when more complicated 
spaces involving the exceptional groups, such as $E_6/(SO(10)\times U(1))$ and $E_7/(E_6 \times U(1))$, are 
considered. For this reason, another way of deriving the (co)tangent bundle was elaborated in \cite{AKL2}, which is 
based on the property of the Hermitian symmetric spaces and supersymmetry considerations. It allows one to obtain
the closed form expression of the tangent bundle Lagrangian. Using this expression, the tangent bundle over 
$E_6/(SO(10)\times U(1))$ was obtained. The cotangent bundle was also obtained with the use of the Legendre 
transformation. The method in \cite{AKL2} can be used for the case of the $E_7/(E_6\times U(1))$ as well, 
 but the Legendre transformation to the cotangent bundle is very cumbersome. Meanwhile the 
closed form expression of the cotangent bundle over any Hermitian symmetric space was obtained in \cite{KuNo}.

The purpose of the paper is to derive the tangent and cotangent bundles over the Hermitian symmetric space 
$E_7/(E_6\times U(1))$.  
We use closed expressions given in \cite{AKL2} and \cite{KuNo} for the tangent bundle 
and the cotangent bundle actions over any Hermitian symmetric space, respectively. The latter one in \cite{KuNo} was presented 
in a matrix form. We first elaborate on this result in order to provide a new closed formula for cotangent 
bundle action, more suitable for calculations. This reformulation will indicate a relation between the K\"ahler potential 
and the cotangent bundle action for any Hermitian symmetric space. More precisely, it will turn out that
pole structure of derivatives of the K\"ahler potential under uniform rescaling of the coordinates completely 
determines the cotangent bundle action.
It was also pointed out in \cite{Ku-Co} that there is a relationship between the tangent bundle action and 
the K\"ahler potential. Our reformulation also shows the existence of such a relation for the cotangent bundle action.

The paper is organized as follows. In Section \ref{review} we give the background material on the $\cN=2$ sigma model in 
projective superspace. The tangent bundle construction in \cite{AKL2} and \cite{KuNo} is briefly reviewed in Section 
\ref{tangent}. We also provide the relationship of the two results. In Section \ref{cotangent}, we explain the cotangent
bundle construction in \cite{KuNo} and reformulate their result for application to $E_7/(E_6\times U(1))$. 
In Section \ref{result}, we derive the tangent and cotangent bundles over $E_7/(E_6\times U(1))$ with the use of the 
results in Sections \ref{tangent} and \ref{cotangent}. Finally, in Section \ref{general} we discuss the relation between the 
K\"ahler potential and the cotangent bundle action for all the Hermitian symmetric spaces.
In the Appendix, to illustrate the power of our approach we 
provide a detailed calculation of the tangent and cotangent bundle action for $E_6/(SO(10)\times U(1))$ and show the 
equivalence of our result to those in \cite{AKL2}.

%%%%%%%%%%%%%%%%%%%%%%%%%%%%%%%%%%%%%%%%%%

\sect{$\cN=2$ Sigma Models in the Projective Superspace}\label{review}
We start with a family of four-dimensional $\cN=2$ off-shell supersymmetric nonlinear sigma models that are described in 
ordinary $\cN=1$ superspace by the action\footnote{The study of such models in this context was initiated in
\cite{GK1,GK2, K}. They correspond to a subclass of the general hypermultiplet theories in projective superspace 
\cite{LR1,LR2}.}
\bea
S[\U, \breve{\U}]  =  
\frac{1}{2\pi {\rm i}} \, \oint \frac{{\rm d}\z}{\z} \,  
 \int {\rm d}^8 z \, 
K \big( \U^I (\z), \breve{\U}^{\bar{J}} (\z)  \big) \,,\quad z^M=(x^\mu,\theta_\alpha,\bar{\theta}^{\dot{\alpha}}),
\label{nact} 
\eea
where $\mu=0,1,2,3$ and $\alpha, \dot{\alpha}=1,2$.
The action is formulated in terms of the so-called polar multiplet \cite{LR1,LR2} (see also \cite{G-RLRWvU}), 
one of interesting $\cN=2$  multiplets living in the projective superspace.
The polar multiplet is described by the so-called arctic superfield $\U(\z)$ and antarctic superfield $\breve{\U} (\z) $ 
that are generated by an infinite set of ordinary $\cN=1$ superfields:
\be
 \U (\z) = \sum_{n=0}^{\infty}  \, \U_n \z^n = 
\F + \S \,\z+ O(\z^2) ~,\qquad
\breve{\U} (\z) = \sum_{n=0}^{\infty}  \, {\bar
\U}_n
 (-\z)^{-n}\,,
\label{exp}
\ee
where $\breve{\U}$ is the conjugation of $\U$ under the composition of complex conjugation with the antipodal map
on the Riemann sphere, $\bar{\zeta}\rightarrow -1/\zeta$. Here $\F$ is chiral, $\S$  complex linear, 
\be
{\bar D}_{\dt{\a}} \F =0~, \qquad \qquad {\bar D}^2 \S = 0 \,,
\label{chiral+linear}
\ee
and the remaining component superfields are unconstrained complex superfields.  
The above theory is obtained as a minimal $\cN=2$ extension of the
general four-dimensional $\cN=1$ supersymmetric nonlinear sigma model \cite{Zumino}
\be
S[\F, \bar \F] =  \int {\rm d}^8 z \, K(\Phi^{I},
 {\bar \Phi}{}^{\bar{J}})  \,,
\label{nact4}
\ee
where $K$ is the K\"ahler potential of a K\"ahler manifold $\cM$.

The extended supersymmetric  sigma model  (\ref{nact}) inherits  all the geometric features of its $\cN=1$ predecessor 
(\ref{nact4}), such as the K\"ahler invariance and invariance under the holomorphic reparametrization of the K\"ahler manifold.
The latter property implies that the variables $(\F^I, \S^J)$ parametrize the tangent bundle $T\cM$ of the K\"ahler 
manifold $\cM$ \cite{K}. 

To describe the theory in terms of the physical superfields $\F$ and $\S$ only, all the auxiliary superfields have to be 
eliminated with the aid of the corresponding algebraic equations of motion
\bea
\oint \frac{{\rm d} \z}{\z} \,\z^n \, \frac{\pa K(\U, \breve{\U} 
) }{\pa \U^I} ~ = ~ \oint \frac{{\rm d} \z}{\z} \,\z^{-n} \, \frac{\pa 
K(\U, \breve{\U} ) } {\pa \breve{\U}^{\bar I} } ~ = ~
0 ~, \qquad n \geq 2 \,.               
\label{asfem}
\eea
Let $\U_*(\z) \equiv \U_*( \z; \F, {\bar \F}, \S, \bar \S )$ denote a unique solution subject to the initial conditions
\bea
\U_* (0)  = \F ~,\qquad  \quad \dt{\U}_* (0) 
 = \S \,.
\label{geo3} 
\eea

For a general K\"ahler manifold $\cM$, the auxiliary superfields $\U_2, \U_3, \dots$, as well as their
conjugates,  can be eliminated  only perturbatively. 
Their elimination  can be carried out, using the ansatz \cite{KL}
\bea
\U^I_n = 
\sum_{p=0}^{\infty} 
G^I{}_{J_1 \dots J_{n+p} \, \bar{L}_1 \dots  \bar{L}_p} (\F, {\bar \F})\,
\S^{J_1} \dots \S^{J_{n+p}} \,
{\bar \S}^{ {\bar L}_1 } \dots {\bar \S}^{ {\bar L}_p }\,, 
\qquad n\geq 2\,.
\eea
Upon elimination of the auxiliary superfields, the action (\ref{nact}) should take the form \cite{GK1,GK2}
\bea
S_{{\rm tb}}[\F,  \S]  
&=& \int {\rm d}^8 z \, \Big\{\,
K \big( \F, \bar{\F} \big)+ \cL\big(\F, \bar \F, \S , \bar \S \big)
\Big\}\,, \label{act-tab}
\eea
where
\bea
\cL\big(\F, \bar \F, \S , \bar \S \big)
&=& \sum_{n=1}^{\infty} \cL^{(n)}
\big(\F, \bar \F, \S , \bar \S \big) \label{tan}\\
&=&\sum_{n=1}^{\infty} \cL_{I_1 \cdots I_n {\bar J}_1 \cdots {\bar 
J}_n }  \big( \F, \bar{\F} \big) \S^{I_1} \dots \S^{I_n} 
{\bar \S}^{ {\bar J}_1 } \dots {\bar \S}^{ {\bar J}_n }\,. \non
\eea
The first term in the expansion (\ref{tan}) is given as $\cL_{I {\bar J} }=  - g_{I \bar{J}} \big( \F, \bar{\F}  \big) $ 
and the tensors $\cL_{I_1 \cdots I_n {\bar J}_1 \cdots {\bar J}_n }$ for  $n>1$ 
are functions of the Riemann curvature $R_{I {\bar J} K {\bar L}} \big( \F, \bar{\F} \big) $ and its covariant 
derivatives.  
Eq. (\ref{act-tab}) is written by the base manifold coordinate $\Phi$ and the tangent space coordinate 
$\Sigma$. Therefore, this is the tangent bundle action.

The complex linear tangent variables $\S$'s in (\ref{act-tab}) can be dualized 
into chiral one-forms in accordance with the generalized Legendre transform \cite{LR1}. 
To construct a dual formulation, we consider the first-order action
\bea
S=  \int {\rm d}^8 z \, \Big\{\,
K \big( \F, \bar{\F} \big)+  \cL
\big(\F, \bar \F, \S , \bar \S \big)
+\J_I \,\S^I + {\bar \J}_{\bar I} {\bar \S}^{\bar I} 
\Big\}\,,
\label{f-o}
\eea
where the tangent vector $\S^I$ is now a complex unconstrained superfield, 
while the one-form $\Psi$ is chiral, ${\bar D}_{\dt \a} \J_I =0$.
Integrating out $\Sigma$'s and their conjugate should give the cotangent-bundle action
\bea
S_{{\rm ctb}}[\F,  \J]  
&=& \int {\rm d}^8 z \, \Big\{\,
K \big( \F, \bar{\F} \big)+    
\cH \big(\F, \bar \F, \J , \bar \J \big)\Big\}\,,
\label{act-ctb}
\eea
where 
\bea
\cH \big(\F, \bar \F, \J , \bar \J \big)&=& \sum_{n=1}^\infty \cH^{(n)}(\Phi,\bar{\Phi},\Psi, \bar{\Psi}) \nonumber \\
&=&\sum_{n=1}^{\infty} \cH^{I_1 \cdots I_n {\bar J}_1 \cdots {\bar 
J}_n }  \big( \F, \bar{\F} \big) \J_{I_1} \dots \J_{I_n} 
{\bar \J}_{ {\bar J}_1 } \dots {\bar \J}_{ {\bar J}_n }\,, \label{h}
\eea
with $\cH^{I {\bar J}} \big( \F, \bar{\F} \big)=g^{I {\bar J}} \big( \F, \bar{\F} \big)$.
The variables $(\Phi^I, \Psi_J)$ parametrize the cotangent bundle $T^*\cM$ of the
K\"ahler manifold $\cM$. The action (\ref{act-ctb}) is written by chiral superfields only,
so the action gives the hyperk\"ahler potential.

Above, we have given a brief sketch of a derivation of the tangent and cotangent bundle actions 
(\ref{act-tab}) and (\ref{act-ctb}).
Detailed calculations of $\cL(\F, \bar \F, \S , \bar \S)$ and $\cH(\F, \bar \F, \J , \bar \J)$ for the case 
when $\cM$ is a Hermitian symmetric space are given in \cite{GK1, GK2, ArNi, AKL, AKL2, KuNo}.
For our purpose the most useful formulation is to use the method as presented in \cite{KuNo}. 
Below, we briefly review the result in \cite{KuNo} and, in particular, reformulate the result of the cotangent 
bundle to be better suited for our purpose.

%%%%%%%%%%%%%%%%%%%%%%%%%%%%%%%%%%%%%%%%%%%%%%%%
\sect{The Tangent Bundle}\label{tangent}

In what follows, we consider the case when $\cM$ is a Hermitian symmetric space:
\be
\nabla_L  R_{I_1 {\bar  J}_1 I_2 {\bar J}_2}
= {\bar \nabla}_{\bar L} R_{I_1 {\bar  J}_1 I_2 {\bar J}_2} =0\,.
\label{covar-const}
\ee
Then, the algebraic equations of motion (\ref{asfem}) are known to be equivalent to the holomorphic 
geodesic equation (with complex evolution parameter) \cite{GK1,GK2}
\bea
\frac{ {\rm d}^2 \U^I_* (\z) }{ {\rm d} \z^2 } + 
\G^I_{~JK} \Big( \U_* (\z), \bar{\F} \Big)\,
\frac{ {\rm d} \U^J_* (\z) }{ {\rm d} \z } \,
\frac{ {\rm d} \U^K_* (\z) }{ {\rm d} \z }  =0 \,,
\eea
under the same initial conditions (\ref{geo3}).
Here $\G^I_{~JK}(\F , \bar{\F} )$ are the Christoffel symbols for the K\"ahler metric $g_{I \bar J} ( \F , \bar{\F} ) = \pa_I 
\pa_ {\bar J}K ( \F , \bar{\F} )$. In particular, we have 
\bea
\U^I_2 = -\hf \G^I_{~JK} \big( \F, \bar{\F} \big) \, \S^J\S^K\,.
\label{U2}
\eea

According to the principles of projective superspace \cite{LR1,LR2},  the action  (\ref{nact}) 
is invariant under the $\cN=2$ supersymmetry transformations
\be
\d \U^I (\z)= {\rm i} \left(\ve^\a_i Q^i_\a +
{\bar \ve}^i_\ad {\bar Q}^\ad_i \right)  \U^I(\z),
\label{SUSY1}
\ee
when $\U^I(\z)$ is viewed as a $\cN=2$ superfield. Here $i=1,2$ is the $SU(2)_R$ index.
However, since the action is given in $\cN=1$ superspace,
it is only the $\cN=1$ supersymmetry which is manifestly 
realized. The second hidden supersymmetry can be shown to act on the 
physical superfields $\F$ and $\S$ as follows (see, e.g. \cite{G-RLRWvU})
\bea
\d \F^I = {\bar \ve}_{\dt{\a}} {\bar D}^{\dt{\a}} \S^I\,, \qquad 
\d \S^I = -\ve^\a D_\a \F^I +   {\bar \ve}_{\dt{\a}} {\bar D}^{\dt{\a}} \U_2^I\,.
\eea 
Upon elimination of the auxiliary superfields, 
the action (\ref{act-tab}), which is  associated with the Hermitian symmetric space 
$\cM$, is invariant under 
\bea
\d \F^I = {\bar \ve}_{\dt{\a}} {\bar D}^{\dt{\a}} \S^I\,, \qquad 
\d \S^I = -\ve^\a D_\a \F^I -\hf   {\bar \ve}_{\dt{\a}} {\bar D}^{\dt{\a}} 
\Big\{ \G^I_{~JK} \big( \F, \bar{\F} \big) \, \S^J\S^K \Big\} \,.
\label{SUSY2}
\eea 

The action  (\ref{act-tab}) varies under (\ref{SUSY2}) as follows
\bea
\d S_{{\rm tb}}[\F,  \S]  
&=& \int {\rm d}^8 z \, \Big\{ 
\frac{\pa \cL}{\pa \F^I} -\frac{\pa \cL}{\pa \S^K}\, \G^K_{~IJ} \S^J\Big\}
{\bar \ve}_{\dt{\a}} {\bar D}^{\dt{\a}} \S^I \non \\
&-& \int {\rm d}^8 z \, \Big\{ 
\hf R_{K {\bar J} L }{}^I\, \frac{\pa \cL}{\pa \S^I}\, \S^K \S^L
+ \frac{\pa \cL}{\pa {\bar \S}^{\bar J} }
+g_{I \bar{J}}\, \S^I \Big\} 
{\bar \ve}_{\dt{\a}} {\bar D}^{\dt{\a}} {\bar \F}^{\bar J}~+~{\rm c.c.}
\eea
Here the variation in the first line vanishes, since the curvature is covariantly constant.
Thus, we obtain the first-order differential equation 
\bea
\hf R_{K {\bar J} L }{}^I\, \frac{\pa \cL}{\pa \S^I}\, \S^K \S^L
+ \frac{\pa \cL}{\pa {\bar \S}^{\bar J} }
+g_{I \bar{J}}\, \S^I =0\,. \label{fd2}
\eea
The solution to the equation \refer{fd2} was first presented in \cite{AKL2} in the form
\bea
 \oper{L} = -\frac{\Exp{R_{\Sigma,\bar\Sigma}}-1}{R_{\Sigma,\bar\Sigma}}\abs{\Sigma}^2, \hspace{1cm} \abs{\Sigma}^2 := g_{I\bar J}\Sigma^{I}\bar\Sigma^{\bar J}\,. \label{tbsol}
\eea
Then an alternative form of the solution was obtained in \cite{KuNo}. In the rest of this section we show that 
the two results are equivalent. 
The solution in \cite{KuNo} is given by
\begin{equation}\label{master}
\oper{L}(\Phi,\bar\Phi,\Sigma,\bar\Sigma) = -\frac{1}{2}\mathbf{\Sigma}^{T}\mathbf{g}\frac{\ln(\mathbf{1}+\mathbf{R}_{\Sigma,\bar\Sigma})}
{\mathbf{R}_{\Sigma,\bar\Sigma}}\mathbf{\Sigma}\,,
\end{equation}
where
\begin{equation}
\mathbf{\Sigma} = \begin{pmatrix}\Sigma^I \\ \bar\Sigma^{\bar I}\end{pmatrix}\,, \hspace{1cm}
\mathbf{g} = \begin{pmatrix}
0 & g_{I\bar J} \\
g_{\bar I J} & 0
\end{pmatrix}\,,
\end{equation}
\begin{equation}
\mathbf{R}_{\Sigma,\bar\Sigma} = \begin{pmatrix}
0 & (R_{\Sigma})_{\phantom{I}\bar J}^{I} \\
(R_{\bar \Sigma})_{\phantom{\bar I}J}^{\bar I}
\end{pmatrix} = \begin{pmatrix}
0 & \frac{1}{2}R_{K\phantom{I}L\bar J}^{\phantom{K}I}\Sigma^{K}\Sigma^{L} \\
\frac{1}{2}R_{\bar K\phantom{\bar I}\bar L J}^{\phantom{\bar K}\bar I}\bar \Sigma^{\bar K}\bar \Sigma^{\bar L}
\end{pmatrix}\,.
\end{equation}
We shall rewrite  (\ref{master}), using the following differential operators
\begin{equation}\label{R}
R_{\Sigma,\bar\Sigma} =-(R_{\Sigma})_{\phantom{I}\bar J}^{I}\bar\Sigma^{\bar J}\frac{\partial}{\partial \Sigma^{I}}, \qquad
\bar R_{\Sigma,\bar \Sigma}=-(R_{\bar \Sigma})_{\phantom{\bar I}J}^{\bar I}\Sigma^{ J}\frac{\partial}{\partial \bar \Sigma^{\bar I}}\,.
\end{equation}
These operators were originally introduced in \cite{AKL2}. They satisfy the relation
\be
R_{\Sigma,\bar\Sigma}{\cal L}^{(n)}=\bar R_{\Sigma,\bar \Sigma}{\cal L}^{(n)}\,.
\ee
Performing the Taylor expansion for (\ref{master}), we have
\begin{equation}\label{taylor}
\oper{L} = \sum_{n=1}^{\infty}c_n F^{(n)}\,, \hspace{1cm} c_n = \frac{(-1)^{n}}{n}\,,
\end{equation}
where the functions $F^{(n)}$ given by
\bea
F^{(2k+2)} &=& \Sigma^{I}g_{I\bar J} \Bigl(\Bigl(R_{\bar\Sigma}R_{\Sigma}\Bigr)^{k}\Bigr)_{\phantom{\bar J}\bar K}^{\bar J}
(R_{\bar\Sigma})_{\phantom{\bar K}L}^{\bar K}\Sigma^{L}\,,\quad k = 0, 1, 2 \ldots \\
F^{(2k+1)} &=& \Sigma^{I}g_{I\bar J} \Bigl(\Bigl(R_{\bar\Sigma}R_{\Sigma}\Bigr)^k\Bigr)_{\phantom{\bar J}\bar K}^{\bar J}\bar\Sigma^{\bar K}\,, \quad k = 0, 1, 2 \ldots 
\eea
satisfy
relations
\begin{eqnarray}
&\displaystyle\Sigma^{I}\frac{\partial}{\partial\Sigma^{I}}F^{(n)} = \bar \Sigma^{\bar I}\frac{\partial}{\partial\bar \Sigma^{\bar I}}F^{(n)} = nF^{(n)}\,,& \\
&\displaystyle{\partial \over \partial \Sigma^I}F^{(2k+2)}=(2k+2)g_{I\bar{J}}\left((R_{\bar{\Sigma}}R_{\Sigma})^kR_{\bar{\Sigma}}\right)^{\bar{J}}_{~K}\Sigma^K,& \\
&\displaystyle{\partial \over \partial \Sigma^I}F^{(2k+1)}=(2k+1)g_{I\bar{J}}\left((R_{\bar{\Sigma}}R_{\Sigma})^k\right)^{\bar{J}}_{~\bar{K}}\bar{\Sigma}^{\bar{K}}.& 
\end{eqnarray}
Using these identities, one can easily prove
\begin{equation}
F^{(n+1)} = \frac{(-R_{\Sigma,\bar\Sigma})^n}{n!}\abs{\Sigma}^2\,.
\end{equation}
Putting this into \refer{taylor}, we find out that
\begin{equation}
\oper{L} = -\sum_{n=1}^{\infty}\frac{(R_{\Sigma,\bar\Sigma})^{n-1}}{n!}\abs{\Sigma}^2 = -\int\limits_{0}^1\diff t\, \Exp{tR_{\Sigma,\bar\Sigma}}
\abs{\Sigma}^2 = -\frac{\Exp{R_{\Sigma,\bar\Sigma}}-1}{R_{\Sigma,\bar\Sigma}}\abs{\Sigma}^2\,. \label{tan-gen}
\end{equation}
This form coincides with (\ref{tbsol}).

\section{The Cotangent Bundle}\label{cotangent}
We start with (\ref{f-o}) to construct a dual formulation \cite{AKL2}. 
The action can be shown to be invariant under the following supersymmetry transformations
\bea
\d \F^I &=&\hf {\bar D}^2 \big\{ \overline{\ve \q} \, \S^I\big\} \,, \non \\
\d \S^I &=& -\ve^\a D_\a \F^I -\hf   {\bar \ve}_{\dt{\a}} {\bar D}^{\dt{\a}} 
\Big\{ \G^I_{~JK} \big( \F, \bar{\F} \big) \, \S^J\S^K \Big\} 
-\hf  \overline{\ve \q} \, \G^I_{~JK} \big( \F, \bar{\F})  \, \S^J {\bar D}^2\S^K 
\,, \non \\
\d \J_I &=&- \hf {\bar D}^2 \Big\{ \overline{\ve \q} \, K_I \big( \F, \bar{\F})  \Big\}
+\hf {\bar D}^2 \Big\{ \overline{\ve \q} \, \G^K_{~IJ} \big( \F, \bar{\F} \big)\,
\S^J \Big\} \,\J_K
\,.
\label{SUSY3}
\eea
These transformations induce the supersymmetry transformations under which the action
(\ref{act-ctb}) is invariant
\bea
\d \F^I &=&\hf {\bar D}^2 \big\{ \overline{\ve \q} \, \S^I  \big(\F, \bar \F, \J , \bar \J \big) \big\} \,, \non \\
\d \J_I &=&- \hf {\bar D}^2 \Big\{ \overline{\ve \q} \, K_I \big( \F, \bar{\F})  \Big\}
+\hf {\bar D}^2 \Big\{ \overline{\ve \q} \, \G^K_{~IJ} \big( \F, \bar{\F} \big)\,
\S^J  \big(\F, \bar \F, \J , \bar \J \big) \Big\} \,\J_K
\,,
\label{SUSY4}
\eea
with 
\bea
\S^I  \big(\F, \bar \F, \J , \bar \J \big) 
= \frac{\pa}{\pa \J_I} \, \cH  \big(\F, \bar \F, \J , \bar \J \big)\equiv \cH^I \,.
\eea
The requirement of invariance under such transformations 
can be shown to be equivalent to the following nonlinear equation 
on $\cH$ \cite{AKL2}:
\be
\cH^I \,  g_{I {\bar J}} - \hf \, \cH^K\cH^L \,  R_{K {\bar J} L}{}^I \,\J_I =
{\bar \J}_{ \bar J} \,. \label{cot-eq}
\ee
This equation also results directly from (\ref{fd2}),  using the definition of the 
$\J$'s, or, if one wants, as a consequence of the  superspace Legendre transform.
Detailed discussion of the above is given in \cite{AKL2}.

The closed form expression of $\cH$ for any Hermitian symmetric space was obtained as a solution of (\ref{cot-eq}), 
which is given by \cite{KuNo}
\begin{equation}\label{master2}
\oper{H}(\Phi,\bar\Phi,\Psi,\bar\Psi) = \frac{1}{2} \mathbf{\Psi}^T\mathbf{g}^{-1}\oper{F}(-\mathbf{R}_{\Psi,\bar\Psi})\mathbf{\Psi}\,,
\end{equation}
where
\begin{equation}
\mathbf{\Psi} = \begin{pmatrix}\Psi_I \\ \bar\Psi_{\bar I}\end{pmatrix}\,, \hspace{1cm}
\mathbf{g}^{-1} = \begin{pmatrix}
0 & g^{I\bar J} \\
g^{\bar I J} & 0
\end{pmatrix}\,,
\end{equation}
\begin{equation}
\mathbf{R}_{\Psi,\bar\Psi} = \begin{pmatrix}
0 & (R_{\Psi})_{I}^{\phantom{I}\bar J} \\
(R_{\bar \Psi})_{\bar I}^{\phantom{\bar I}J} & 0
\end{pmatrix} = \begin{pmatrix}
0 & \frac{1}{2}R_{I}^{\phantom{I}K\bar J L}\Psi_{K}\Psi_{L} \\
\frac{1}{2}R_{\bar I}^{\phantom{\bar I}\bar K J\bar L }\bar \Psi_{\bar K}\bar \Psi_{\bar L} & 0
\end{pmatrix}\,,
\end{equation}
and
\begin{equation}
\oper{F}(x) = \frac{1}{x}\biggl[\sqrt{1+4x}-1-\ln\biggl(\frac{1+\sqrt{1+4x}}{2}\biggr)\biggr]\,.
\end{equation}

We rewrite (\ref{master2}) to a convenient form for our purpose.
In a similar way to the tangent bundle case, we introduce the following differential operators
\begin{equation}
R_{\Psi,\bar\Psi} = -(R_{\Psi})_{I}^{\phantom{I}\bar J}\bar\Psi_{\bar J}\frac{\partial}{\partial\Psi_I} \,,\quad 
\bar R_{\Psi,\bar\Psi}=-(R_{\bar\Psi})_{\bar I}^{\phantom{\bar I}J}\Psi_{ J}\frac{\partial}{\partial\bar\Psi_{\bar I}}\,. \label{diff}
\end{equation}
They satisfy
\be
R_{\Psi,\bar\Psi}\cH^{(n)}=\bar R_{\Psi,\bar\Psi} \cH^{(n)}.
\ee
Performing the Taylor expansion for \refer{master2}, we have
\begin{equation}
\oper{H} = \sum_{n=1}^{\infty}c_nG^{(n)}\,, \hspace{1cm} c_n = \frac{(-1)^{n-1}\oper{F}^{(n-1)}(0)}{(n-1)!}\,, \label{H1}
\end{equation}
where the terms  $G^{(n)}$ are given by
\bea
G^{(2k+2)} &=& \Psi_{I}g^{I\bar J} \Bigl(\Bigl(R_{\bar\Psi}R_{\Psi}\Bigr)^{k}\Bigr)_{\bar J}^{\phantom{\bar J}\bar K}
(R_{\bar\Psi})_{\bar{K}}^{\phantom{\bar K}L}\Psi_{L}\,, \quad k= 0, 1, 2 \ldots \\
G^{(2k+1)} &=& \Psi_{I}g^{I\bar J} \Bigl(\Bigl(R_{\bar\Psi}R_{\Psi}\Bigr)^k\Bigr)_{\bar{J}}^{\phantom{\bar J}\bar K}\bar\Psi_{\bar K}\,,
\quad k = 0, 1, 2 \ldots 
\eea
and satisfy the following relations
\begin{eqnarray}
&\displaystyle \Psi_{I}\frac{\partial}{\partial\Psi_{I}}G^{(n)} = \bar \Psi_{\bar I}\frac{\partial}{\partial\bar \Psi_{\bar I}}G^{(n)} = nG^{(n)}\,, &\\
&\displaystyle{\partial \over \partial \Psi_I}G^{(2k+2)}=(2k+2)g^{I\bar J} \Bigl(\Bigl(R_{\bar\Psi}R_{\Psi}\Bigr)^{k}\Bigr)_{\bar J}^{\phantom{\bar J}\bar K}(R_{\bar\Psi})_{\bar{K}}^{\phantom{\bar K}L}\Psi_{L}\,,& \\
&\displaystyle{\partial \over \partial \Psi_I}G^{(2k+1)}=(2k+1) g^{I\bar J} \Bigl(\Bigl(R_{\bar\Psi}R_{\Psi}\Bigr)^k\Bigr)_{\bar{J}}^{\phantom{\bar J}\bar K}\bar\Psi_{\bar K}\,.& 
\end{eqnarray}
By using the above identities, one can show that
\begin{equation}\label{G1}
G^{(n+1)} = \frac{(-R_{\Psi,\bar\Psi})^n}{n!}\abs{\Psi}^2\,, \hspace{1cm} \abs{\Psi}^2 := g^{I\bar J}\Psi_{I}\bar\Psi_{\bar J}\,.
\end{equation}
Putting together \refer{H1} and \refer{G1}, we find
\bea
\cH=\sum_{n=0}^\infty {\cF^{(n)}(0) \over n!}\frac{R_{\Psi,\bar\Psi}^{\,n}}{n!}|\Psi|^2\,.
\eea
Now we make use of the formula
\begin{equation}
\frac{x^n}{n!} = \oint_{C}\frac{\diff \xi}{2\pi\iunit} \frac{\Exp{\xi x}}{\xi^{n+1}}\,,
\end{equation}
where, at this point, contour $C$ can be any closed loop containing  the origin with a positive (counterclockwise) orientation.
Using this identity, we obtain
\begin{equation}
\oper{H} = \sum_{n=0}^\infty {\cF^{(n)}(0) \over n!}\oint_C {\diff\xi \over 2\pi \iunit}{e^{\xi R_{\Psi,\bar{\Psi}}} \over \xi^{n+1}}|\Psi|^2\,, \label{gen}
\end{equation}
where now the contour $C$ must be chosen such that it lies in the region where the factor $e^{\xi R_{\Psi,\bar{\Psi}}}|\Psi|^2$ 
is analytic. 
With this in mind, we can rewrite eq. (\ref{gen}) as
\be
\oper{H}=\oint_{C}^{} \frac{\diff \xi}{2\pi\iunit} \frac{\oper{F}(1/\xi)}{\xi}\Exp{\xi R_{\Psi,\bar \Psi}}|\Psi|^2\,. \label{gen2}
\ee
The function $\oper{F}(1/\xi)/\xi$ gives a branch cut between $-4$ and $0$. Therefore, the contour has to be 
chosen so that it does not cross the branch cut, for the $\cH$ is real and regular. Combining both 
requirements, we see that the contour $C$ in addition to avoiding the branch cut must be also bounded by the poles of the factor 
$e^{\xi R_{\Psi,\bar{\Psi}}}|\Psi|^2$ (see fig. \ref{fig1}). In the subsequent section, we derive the tangent bundle as well as 
the cotangent bundle over $E_7/[E_6\times U(1)]$ by using (\ref{tan-gen}) and (\ref{gen2}). 

\begin{figure}
\begin{center}
{\small
\begin{tikzpicture}
\draw[dashed] (-5,0) -- (0,0)  -- (5,0);
\draw[thick,|-|] (-4,0) node[below=2pt]{-4} -- (0,0) node[below right]{0};
\draw[dashed] (0,3) -- (0,-3);
\draw plot[smooth cycle,->] coordinates{(-4.5,0) (-3,2) (-2,0.2) (0,0.5) (1,0) (-2,-2)};
\draw[thick] plot[mark=x,mark size=4pt] coordinates{(-2,0.4)};
\draw[thick] plot[mark=x,mark size=4pt] coordinates{(0.5,2)};
\draw[thick] plot[mark=x,mark size=4pt] coordinates{(-4,3)};
\draw[thick] plot[mark=x,mark size=4pt] coordinates{(1,-2)};
\draw[thick] plot[mark=x,mark size=4pt] coordinates{(2,1)};
\draw (2,1) node[right=2pt]{poles of $\Exp{\xi R_{\Psi,\bar\Psi}}\abs{\Psi}^2$};
\draw (-2,-2) node[below right]{$C$};
\draw (-2,-2) node{$>$};
\end{tikzpicture}
\caption{The contour of integration in eq. \refer{gen2}.}
\label{fig1}}
\end{center}
\end{figure}

%%%%%%%%%%%%%%%%%%%%%%%%%%%%%%%%%%%%%
\sect{The Hermitian Symmetric Space $E_7/(E_6 \times U(1))$}\label{result}
The K\"ahler potential for the Hermitian symmetric space $E_7/(E_6\times U(1))$ was computed in \cite{DV, HiNi}. 
Here we will use the K\"ahler potential derived in \cite{HiNi}. In order to be consistent with the notation adopted in \cite{HiNi}, 
we will use small Latin letters to label indices.  Upper indices will be used for the base-space  ($\Phi^I\rightarrow \Phi^i$) and 
tangent coordinates ($\Sigma^I \rightarrow \Sigma^i$), while lower indices are reserved for one-forms ($\Psi_I\rightarrow \Psi_i$).

Locally, the Hermitian symmetric space $E_7/(E_6\times U(1))$ can be described by complex variables $\F^i$ transforming in 
the ${\bf 27}$ representation of the $E_6$ and their conjugates.
\begin{eqnarray}
 \F^i~,\qquad \bar{\F}_i:=(\F^i)^*~, 
 \qquad \quad
 i=1,\dots,27~.
\end{eqnarray}
The K\"ahler potential is
\begin{eqnarray}
 K(\F,\bar{\F})=\ln\left(1+\F^i \bar{\F}_{i}
  +\frac{1}{4}
   |\Gamma_{ijk}\F^j\F^k|^2
  +\frac{1}{36}
   |\Gamma_{ijk}\F^i\F^j\F^k|^2
  \right)\,.
\end{eqnarray}
Here we have used the notation
$
 |\Gamma_{ijk}\F^j\F^k|^2=
   (\Gamma_{ijk}\F^j\F^k)(\Gamma^{ilm}\bar{\F}_l\bar{\F}_m)
$
and
$
 |\Gamma_{ijk}\F^i\F^j\F^k|^2=
    (\Gamma_{ijk}\F^i\F^j\F^k)(\Gamma^{lmn}\bar{\F}_l\bar{\F}_m\bar{\F}_n)
$
where $\Gamma_{ijk}$ is a rank-3 symmetric tensor covariant under the $E_6$. Its complex conjugate is denoted by $\Gamma^{ijk}$. 

Products of these tensors satisfy the identity \cite{KV}
\begin{eqnarray}
 \G_{ijk}\G^{ijl}=10\d_k^l\,.
\end{eqnarray}
And 
\bea
\Gamma_{ijk}\Bigl(\Gamma^{ilm}\Gamma^{jnp}+\Gamma^{ilp}\Gamma^{jnm}+\Gamma^{iln}\Gamma^{jmp}\Bigr) 
=
\delta_k^l\Gamma^{mnp}+\delta_k^p\Gamma^{mnl}+\delta_k^n\Gamma^{mlp}+\delta_k^m\Gamma^{lnp}\,, \label{springer}
\eea
which is called the Springer relation \cite{Springer}.

Let us  calculate the tangent bundle Lagrangian by using \refer{tan-gen}.
In our notation, the first-order differential operator defined in (\ref{R}) is given by
\begin{eqnarray}
 R_{\S,\bar{\S}}=-{1 \over
  2}\S^i\bar{\S}_j\S^k R_{i~k}^{~j~l}(g^{-1})_l^{~m}\,
 {\partial \over \partial \S^m}\,, \label{Rs}
\end{eqnarray}
where $(g^{-1})_i^{~j}=(g_j^{~i})^{-1}$ is the  inverse metric of $g_{i}^{~j}$, that is $g_{i}^{~k}(g^{-1})_{k}^{~j}=\delta_i^{~j}$.
Since we consider a symmetric space, it is actually sufficient to carry out the calculations  at 
a particular point, say  at $\F=0$. The Riemann tensor at $\F=0$ is given as
\begin{eqnarray}
 R_{i~k}^{~j~l}{\Big |}_{\F=0}
 &=&\partial_k\partial^l g_{i}^{~j}
 -(g^{-1})_m^{~n}\partial_n g_i^{~j}\partial^m g_k^{~l}{\Big
 |}_{\F=0} \nonumber \\
 &=&-\d_i^{~j}\d_k^{~l}+\G_{mik}\G^{mlj}-\d_k^{~j}\d_i^{~l}\,.
\end{eqnarray}
Substituting this into (\ref{Rs}), we have
\begin{equation}
R_{\Sigma,\bar\Sigma}= xD-\frac{1}{2}y\partial_x-\frac{1}{3}z\partial_y\,, \hspace{1cm} D:= x\partial_x + y\partial_y+z\partial_z\,,
\end{equation}
where we have used \refer{springer} and invariant quantities under the $E_6$ action
\begin{align}
x & := \Sigma^i\bar\Sigma_i\,, \\
y & := (\Gamma_{ijk}\Sigma^j\Sigma^k)(\Gamma^{ilm}\bar\Sigma_l\bar\Sigma_m)\,, \\
z & := (\Gamma_{ijk}\Sigma^i\Sigma^j\Sigma^k)(\Gamma^{lmn}\bar\Sigma_l\bar\Sigma_m\bar\Sigma_n)\,.
\end{align}
Using 
the Baker-Campbell-Hausdorff expansion formula, one can show that
\begin{equation}
\Exp{tR_{\Sigma,\bar\Sigma}} = \Exp{txD}\Exp{-\frac{t}{2}y\partial_x}\Exp{-\frac{t}{3}z\partial_y}\Exp{\frac{t^2}{12}z\partial_x}
\Exp{-\frac{t^2}{4}yD}\Exp{-\frac{t^3}{18}zD}\,. \label{BCH-tan}
\end{equation}
By a straightforward application of each exponential we obtain
\begin{equation}
\Exp{tR_{\Sigma,\bar\Sigma}}x = -\partial_t \ln\Omega(t;x,y,z)\,, \hspace{1cm} \Omega(t;x,y,z) := 1-tx+\frac{t^2}{4}y-\frac{t^3}{36}z\,. \label{tR-tan}
\end{equation} 
Plugging \refer{tR-tan} into \refer{tan-gen}, we get the tangent bundle action at $\Phi=0$.
\begin{equation}
\oper{L} = \ln\Bigl(1-x+\frac{1}{4}y-\frac{1}{36}z\Bigr)\,.
\end{equation}
This result can be extended to an arbitrary point $\F$ of the base manifold by replacing
\begin{align}
 x & \rightarrow ~ g_i^{~j}\S^i\bar{\S}_j\,, \\
 {1 \over 4}y  & \rightarrow~ 
   {1 \over 2}(g_i^{~j}\S^i\bar{\S}_j)^2
  +{1 \over 4}R_{i~k}^{~j~l}\S^i\bar{\S}_j\S^k\bar{\S}_l\,,  \\
-{1 \over 36}z & \rightarrow -{1 \over 6}(g_i^{~j}\S^i\bar{\S}_j)^3 
   -{1 \over 4}(g_i^{~j}\S^i\bar{\S}_j)(R_{k~l}^{~m~n}\S^k\bar{\S}_m\S^l\bar{\S}_n) \nonumber \\
  &  ~~~~-{1 \over 12}|g_i^{~j}R_{j~l}^{~k~m}\bar{\S}_k\S^l\bar{\S}_m|^2\,. \label{recover}
\end{align}

Let us turn to the cotangent bundle action. Again we restrict the calculations to the origin of the base manifold $\Phi=0$. 
Defining $E_6$ invariant quantities in terms of the cotangent vector
\begin{align}
\tilde{x} & := \Psi_i\bar\Psi^i\,, \\
\tilde{y} & := (\Gamma^{ijk}\Psi_j\Psi_k)(\Gamma_{ilm}\bar\Psi^l\bar\Psi^m)\,, \\
\tilde{z} & := (\Gamma^{ijk}\Psi_i\Psi_j\Psi_k)(\Gamma_{lmn}\bar\Psi^l\bar\Psi^m\bar\Psi^n)\,,
\end{align}
we obtain the differential operator (\ref{diff}) of the form
\bea
 R_{\Psi,\bar{\Psi}}=\tilde{x}\tilde{D}-{1 \over 2}\tilde{y}\partial_{\tilde{x}}-{1 \over 3}\tilde{z}\partial_{\tilde{y}}\,,
 \quad \tilde{D}:=\tilde{x}\partial_{\tilde{x}}+\tilde{y}\partial_{\tilde{y}}+\tilde{z}\partial_{\tilde{z}}\,.
\eea
Repeating the same calculation as below (\ref{BCH-tan}), we find
\be \label{eqR}
e^{\xi R_{\Psi,\bar{\Psi}}}\tilde x=-\partial_\xi \ln\Omega(\xi;\tilde{x},\tilde{y},\tilde{z})\,, 
\ee
where the function $\Omega$ is given in \refer{tR-tan}.
Eq. (\ref{gen2}) with the above leads to
\begin{equation}
\oper{H} = -\oint_C \frac{\diff \xi}{2\pi\iunit}
\frac{\oper{F}(1/\xi)}{\xi}\frac{-\tilde{x}+\frac{\xi}{2}\tilde{y}-\frac{\xi^2}{12}\tilde{z}}{1-\xi \tilde{x} +\frac{\xi^2}{4}\tilde{y}-\frac{\xi^3}{36}\tilde{z}}\,. \label{int-co}
\end{equation}
Note that the factor in (\ref{gen}) $e^{\xi R_{\Psi,\bar{\Psi}}}|\Psi|^2$ produces poles which are given by the roots of the
cubic equation:
\be 
1-\xi x +\frac{\xi^2}{4}y-\frac{\xi^3}{36}z = \frac{z}{36}(\xi_1-\xi)(\xi_2-\xi)(\xi_3-\xi) = 0\,.
\ee
By construction, these poles are not inside the contour which only encircles the branch cut between the origin $\xi=0$ and $\xi=-4$.
Since the function $\cF({1/\xi})/\xi$ has no singularity at infinity in the $\xi$-plane,
the contour can be equivalently respected as encircling the poles $\xi_1, \xi_2$ and $\xi_3$ in the opposite direction.
Thus, we can apply the Residue theorem, picking additional minus sign from the orientation of contour (which kills another minus given by  eq. \refer{eqR}) to obtain
\begin{equation}\label{e7sol1}
\oper{H} = \frac{\oper{F}(1/\xi_1)}{\xi_1}+\frac{\oper{F}(1/\xi_2)}{\xi_2}+\frac{\oper{F}(1/\xi_3)}{\xi_3}\,.
\end{equation}
The explicit form of $\xi_1, \xi_2$ and $\xi_3$ are given by
\bea \label{e7sol2}
 \xi_1&=& {3\tilde{y} \over \tilde{z}}+{2^{1/3}(-81\tilde{y}^2+108\tilde{x}\tilde{z}) \over 3\tilde{z}(-1458\tilde{y}^3+2916\tilde{x}\tilde{y}\tilde{z}-972\tilde{z}^2+A)^{1/3}} \non \\
         &&~\quad -{(-1458\tilde{y}^3+2916\tilde{x}\tilde{y}\tilde{z}-972\tilde{z}^2+A)^{1/3} \over 3 \cdot 2^{1/3}\tilde{z}}, \\\label{e7sol3}
 \xi_2&=& {3 \tilde{y} \over \tilde{z}} - {(1 + i\sqrt{3}) (-81 \tilde{y}^2 + 108 \tilde{x} \tilde{z}) \over 3\cdot 2^{2/3}
     \tilde{z} (-1458 \tilde{y}^3 + 2916 \tilde{x}\tilde{y}\tilde{z} - 972 \tilde{z}^2 + A^{1/3})} \non \\
&&+ {1 - i\sqrt{3} \over 6\cdot 2^{1/3} \tilde{z}} (-1458 \tilde{y}^3 + 2916 \tilde{x} \tilde{y} \tilde{z} - 972 \tilde{z}^2 
    +  A^{1/3}),\\\label{e7sol4}
 \xi_3&=&{3 \tilde{y} \over \tilde{z}} - {(1 - i\sqrt{3}) (-81 \tilde{y}^2 + 108 \tilde{x} \tilde{z}) \over 3\cdot 2^{2/3}
     \tilde{z} (-1458 \tilde{y}^3 + 2916 \tilde{x}\tilde{y}\tilde{z} - 972 \tilde{z}^2 + A^{1/3})} \non \\
&&+ {1 + i\sqrt{3} \over 6\cdot 2^{1/3} \tilde{z}} (-1458 \tilde{y}^3 + 2916 \tilde{x} \tilde{y} \tilde{z} - 972 \tilde{z}^2 
    +  A^{1/3}),
\eea
where $A=\sqrt{4(-81\tilde{y}^2+108\tilde{x}\tilde{z})^3+(-1458\tilde{y}^3+2916\tilde{x}\tilde{y}\tilde{z}-972\tilde{z}^2)^2}$.
The result at an arbitrary point of $\Phi$ can be obtained by the following replacements
\begin{align}
\tilde{x} & \rightarrow~ (g^{-1})_i^{~j}\Psi_j\bar{\Psi}^i\,, \\
 {1 \over 4}\tilde{y}  & \rightarrow~ 
   {1 \over 2}((g^{-1})_i^{~j}\Psi_j\bar{\Psi}^i)^2
  +{1 \over 4}\tilde{R}_{i~k}^{~j~l}\bar{\Psi}^i\Psi_j\bar{\Psi}^k\Psi_l\,,  \\
 -{1 \over 36}\tilde{z} & \rightarrow -{1 \over 6}((g^{-1})_i^{~j}\Psi_j\bar{\Psi}^j)^3
-{1 \over 4}((g^{-1})_i^{~j}\Psi_j\bar{\Psi}^i)(\tilde{R}_{k~m}^{~l~~n}\bar{\Psi}^k\Psi_l\bar{\Psi}^m\Psi_n)\nonumber \\
 & ~~~~-{1 \over 12}|(g^{-1})_i^{~j}\tilde{R}_{j~l}^{~k~m}\Psi_k\bar{\Psi}^l\Psi_m|^2\,, \label{recover-co}
\end{align}
where $\tilde{R}_{i~k}^{~j~l}=(g^{-1})_i^{~m}(g^{-1})_n^{~j}(g^{-1})_k^{~p}(g^{-1})_q^{~l}R_{m~p}^{~~n~q}$.

\section{Generalization}\label{general}
We have constructed the tangent and cotangent bundles over $E_7/(E_6\times U(1))$ by using the projective 
superspace formalism. Our results have been obtained and based on the results in \cite{AKL2} and \cite{KuNo}, 
which give the closed form expressions of the tangent and cotangent bundles. We have rewritten the expression of
the cotangent bundle to a form convenient for our purpose. The new formula for the cotangent bundle action 
(\ref{gen2}) is useful, for it drastically simplifies calculations and allows us to obtain the cotangent bundle action.

Our reformulation suggests an alternative expression for the cotangent bundle 
over any compact Hermitian symmetric space. In order to explain it, let us look at the equation (\ref{int-co}). 
We can rewrite it as follows
\be
\cH=-\oint_C {\diff\xi \over 2\pi \iunit}{\cF(1/\xi) \over \xi}\partial_\xi \ln\Omega(\xi; \tilde{x},\tilde{y},\tilde{z})\,.
\ee
It is easy to see that $\Omega$ is related to the K\"ahler potential $K(\Phi,\bar\Phi)$ by uniform rescaling of coordinates
\be
\ln\Omega(\xi; \tilde{x},\tilde{y},\tilde{z})=K(\Phi\rightarrow -\sqrt{\xi}\bar\Psi, \bar{\Phi}\rightarrow \sqrt{\xi}\Psi):=K_{\xi}(\Psi, \bar{\Psi})\,.
\ee
Thus, 
we see that the cotangent bundle action for any Hermitian symmetric space is given by
\be
\cH=-\oint_C {\diff\xi \over 2\pi \iunit}{\cF(1/\xi) \over \xi}\partial_\xi K_{\xi}(\Psi, \bar{\Psi})\,, \label{gen-H}
\ee
where, as in the $E_7/(E_6\times U(1))$ case, the contour $C$ must encircle the branch cut of $\oper{F}(1/\xi)/\xi$ and 
it must be bounded by the poles of $\partial_\xi K_\xi$, in the same manner as depicted in fig.~\ref{fig1}.
Since for all Hermitian symmetric spaces the $\Omega$ is a finite-order polynomial in $\xi$, we may write
\be
\Omega \sim \Pi_{i}\bigl(\xi -\xi_i(\Psi,\bar\Psi)\bigr)\,,
\ee
where $\xi_i$ are the solutions to characteristic equation
$\Omega = 0$. Since we are interested in (the derivative of) the logarithm of the above quantity, the constant of proportionality is actually not important.
Thus, we are led to
\be
 \partial_{\xi}K_{\xi} = \sum_{\xi}\frac{1}{\xi-\xi_i}\,.
\ee
Substituting this back into (\ref{gen-H}) and using the Residue theorem, we obtain
\be\label{gen3}
\cH=-\sum_i\oint_C {\diff \xi \over 2\pi \iunit}{\cF(1/\xi) \over \xi}{1 \over \xi-\xi_i}=\sum_{i}{\cF(1/\xi_i) \over \xi_i}\,,
\ee
where minus sign is absorbed because the contour $C$ encircles poles in the clockwise direction.
We checked that the results obtained by this formula for all Hermitian symmetric spaces
perfectly coincide with the previous results in \cite{GK1, GK2, AKL, AKL2}.
In the Appendix, as an illustration, we apply the formula (\ref{gen3}) to $E_6/[SO(10)\times U(1)]$ case and show that 
the result coincides with the one in \cite{AKL}.

\section{Conclusion}

In this paper we have presented closed formulas for tangent and cotangent bundle 
action over Hermitian symmetric space $E_7/(E_6\times U(1))$ for the first time, starting from the result in \cite{KuNo}.
This particular case helped us to establish the correspondence between a cotangent bundle action and poles of 
derivatives of the K\"ahler potential under uniform rescaling of coordinates. Based on this correspondence, we 
have presented a general formula for a cotangent bundle action for any Hermitian symmetric space in Eq. \refer{gen3}.
In the former methods in \cite{GK1, GK2, AKL, AKL2}, the Legendre transform is used to obtain a cotangent 
bundle action. This procedure, although in principle applicable to the $E_7/(E_6\times U(1))$ case, is not 
straightforward and one often must find a suitable  ansatz to solve involved algebraic equations.
Our simple formula for the cotangent bundle action obtained from the result in \cite{KuNo} avoids such
a complication and  clearly demonstrates the reduction of work in constructing a cotangent bundle action.

\noindent
{\bf Acknowledgements:}\\
M.A. thanks Sergei M. Kuzenko for useful discussions and comments at the early stage of this work.
F.B. thanks Petr Blaschke for his comments about the contour integral in the cotangent bundle formulation and Veronika
Porte\v{s}ov\'{a} for her aid in correcting the manuscript.
The work of the authors is supported in part by the Research Program MSM6840770029,
by the project of International Cooperation ATLAS-CERN of the Ministry of Education, Youth 
and Sports of the Czech Republic, and by the Japan Society for the Promotion of Science (JSPS) 
and Academy of Sciences of the Czech Republic (ASCR) under the Japan - Czech Republic 
Research Cooperative Program.

\appendix
\section{Cotangent Bundle over $E_6/SO(10)\times U(1)$}

K\"ahler potential or $E_6/ SO(10)\times U(1)$ is given as
\begin{equation}
K(\Phi,\bar\Phi) = \ln\biggl(1+\Phi_{\alpha}\bar\Phi^{\alpha}+\frac{1}{8}\bigl(\bar\Psi^{\alpha}(\sigma_A)_{\alpha \beta}\bar\Psi^{\beta}\bigr)
\bigl(\Psi_{\gamma}(\sigma_A^{\dagger})^{\gamma \delta}\Psi_{\delta}\bigr)\biggr)\,,
\end{equation}
where $(\sigma_A)_{\alpha \beta} = (\sigma_A)_{\beta \alpha}$ are the $16 \times 16$ sigma-matrices which generate, along with their
Hermitian-conjugates $(\sigma_A^{\dagger})^{\alpha \beta}$, the ten-dimensional Dirac matrices in the Weyl representation. 
The sigma-matrices obey the anti-commutation relations
\begin{equation}
(\sigma_A\sigma_B^{\dagger}+\sigma_B\sigma_A^{\dagger})_{\alpha}^{\phantom{\alpha}\beta} = 2\delta_{AB}\delta_{\alpha}^{\phantom{\alpha}\beta}\,.
\end{equation}

As usual, we limit our discussion only to the base point $\Phi=\bar\Phi =0$, where the metric and Riemann tensor are given as
\begin{align}
g_{\alpha}^{\phantom{\alpha}\beta} & \xrightarrow[\Phi = \bar\Phi = 0]{} \delta_{\alpha}^{\phantom{\alpha}\beta}\,, \\
R_{\phantom{\alpha}\beta\phantom{\gamma}\delta}^{\alpha\phantom{\beta}\gamma} &  \xrightarrow[\Phi = \bar\Phi = 0]{}
-\delta_{\beta}^{\phantom{\beta}\alpha}\delta_{\delta}^{\phantom{\delta}\gamma}-
\delta_{\delta}^{\phantom{\delta}\alpha}\delta_{\beta}^{\phantom{\beta}\gamma}
+\frac{1}{2}(\sigma_A)_{\beta\delta}(\sigma_A^{\dagger})^{\alpha\gamma}\,.
\end{align}
In this limit, the operator $R_{\Sigma,\bar\Sigma}$ \refer{R} can be written in the form
\begin{equation}
R_{\Sigma,\bar\Sigma} \xrightarrow[\Phi = \bar\Phi = 0]{} 
\Sigma_{\alpha}\bar\Sigma^{\alpha}\Sigma_{\beta}\frac{\partial}{\partial\Sigma_{\beta}}
-\frac{1}{4}\bigl(\Sigma_{\alpha}(\sigma_A^{\dagger})^{\alpha \beta}\Sigma_{\beta}\bigr)(\sigma_A)_{\gamma \delta}\bar\Sigma^{\gamma}\frac{\partial}{\partial\Sigma_{\delta}}.
\end{equation}
As before, we use the coordinates given by two algebraically independent invariants
\begin{equation}
x := \Sigma_{\alpha}\bar\Sigma^{\alpha}\,, \hspace{1cm} y := \bigl(\Sigma_{\alpha}(\sigma_A^{\dagger})^{\alpha \beta}\Sigma_{\beta}\bigr)
\bigl(\bar\Sigma^{\gamma}(\sigma_A)_{\gamma \delta}\bar\Sigma^{\delta}\bigr)\,, 
\end{equation}
so that the operator $R_{\Sigma,\bar\Sigma}$ becomes
\begin{equation}
R_{\Sigma,\bar\Sigma} = xD -\frac{1}{4}y\partial_x\,, \hspace{1cm} D:= x\partial_x + y\partial_y\,.
\end{equation}
Using  the Baker--Campbell--Hausdorf's formula, 
one finds that
\begin{gather}
\Exp{tR_{\Sigma,\bar\Sigma}} = \Exp{txD}\Exp{-\tfrac{ty}{4}\partial_x}\Exp{-\tfrac{t^2y}{8}D}\,, \\
\Exp{tR_{\Sigma,\bar\Sigma}} x = -\partial_t \ln\Bigl(1-tx+t^2\tfrac{y}{8}\Bigr)\,.
\end{gather}
Plugging this into the formula \refer{tan-gen}, we obtain the tangent bundle action
\begin{equation}
\oper{L} \xrightarrow[\Phi = \bar\Phi = 0]{} \ln\Bigl(1-x+\tfrac{y}{8}\Bigr)\,,
\end{equation}
which is in a complete agreement with the result given in \cite{AKL2}. 

The cotangent bundle action of $E_6/(SO(10)\times U(1))$ can be computed by a direct use of \refer{gen3} after finding the roots of
\begin{equation}
1-\xi x+\xi^2\frac{y}{8} = 0\,. 
\end{equation}
Those are
\begin{equation}
\xi_1 = \frac{4x}{y}-\frac{2\sqrt{4x^2-2y}}{y}\,, \hspace{1cm} \xi_2 = \frac{4x}{y}+\frac{2\sqrt{4x^2-2y}}{y}\,.
\end{equation}
Thus
\begin{equation}
\oper{H} = {\oper{F}(1/\xi_1) \over \xi_1}+{\oper{F}(1/\xi_2)\over \xi_2}\,.
\end{equation}

The result  given in \cite{AKL2} at the base point $\Phi = \bar\Phi = 0$ is formulated as
\begin{equation}
\oper{H} = -\ln\Bigl(\Lambda +\sqrt{\Lambda+x}\Bigr)+\Lambda+\sqrt{\Lambda+x}-\frac{1}{2}\frac{y}{\lambda+\sqrt{\Lambda+x}}\,,
\end{equation}
where
\begin{equation}
\Lambda := \frac{1}{2}+\sqrt{\frac{1}{4}+x+\frac{y}{2}}\,.
\end{equation}
It can be easily proved that our results are in complete agreement with this one up to the irrelevant 
constant factor $\ln2-2$.

\small{

}

\begin{thebibliography}{66}

\bibitem{Zumino}
%
  B.~Zumino,
  ``Supersymmetry and K\"ahler manifolds,''
  Phys.\ Lett.\ B {\bf 87} (1979)  203.
%
\bibitem{Alvarez-Gaume:1981hm}
  L.~Alvarez-Gaum\'e and D.~Z.~Freedman,
  ``Geometrical structure and ultraviolet finiteness in 
  the supersymmetric sigma model,''
  Commun.\ Math.\ Phys.\  {\bf 80} (1981) 443.
%
\bibitem{Karlhede:1984vr}
  A.~Karlhede, U.~Lindstr\"om and M.~Ro\v cek,
  ``Self-interacting tensor multiplets in N=2 superspace,''
  Phys.\ Lett.\ B {\bf 147} (1984) 297.
%
\bibitem{GaIvKaOgSo}
  A.~Galperin, E.~Ivanov, S.~Kalitsyn, V.~Ogievetsky and E.~Sokatchev,
  ``Unconstrained N=2 Matter, Yang-Mills and Supergravity Theories in Harmonic Superspace,''
  Class.\ Quant.\ Grav.\  {\bf 1} (1984) 469;
% \bibitem{Galperin:2001uw}
  A.~S.~Galperin, E.~A.~Ivanov, V.~I.~Ogievetsky and E.~S.~Sokatchev,
  ``Harmonic superspace,''
  Cambridge, UK: Univ. Pr. (2001) 306 p.
%
%\textcolor{red}{\bibitem{Lindstrom:1983rt}
%  U.~Lindstr\"om and M.~Ro\v{c}ek,
%  ``Scalar tensor duality and N=1, N=2 nonlinear sigma models,''
%  Nucl.\ Phys.\ B {\bf 222} (1983) 285.}
%
\bibitem{LR1}
  U.~Lindstr\"om and M.~Ro\v{c}ek,
  ``New hyperk\"ahler  metrics  and new supermultiplets,''
  Commun.\ Math.\ Phys.\  {\bf 115} (1988)  21.
%
\bibitem{LR2}
  U.~Lindstr\"om and M.~Ro\v{c}ek,
  ``N=2 super Yang-Mills theory in projective superspace,''
  Commun.\ Math.\ Phys.\  {\bf 128} (1990) 191.
%
\bibitem{G-RLRWvU}
  F.~Gonzalez-Rey, U.~Lindstr\"om, M.~Ro\v{c}ek, S.~Wiles and R.~von Unge,
  ``Feynman rules in N = 2 projective superspace. I: Massless
  hypermultiplets,''
  Nucl.\ Phys.\  B {\bf 516} (1998) 426 
  [hep-th/9710250].
%
%\bibitem{Hitchin:1986ea}
%  N.~J.~Hitchin, A.~Karlhede, U.~Lindstr\"om and M.~Ro\v cek,
%  ``Hyperk\"ahler metrics and supersymmetry,''
%  Commun.\ Math.\ Phys.\  {\bf 108}, 535 (1987).
%
\bibitem{GK1}
  S.~J.~Gates Jr. and S.~M.~Kuzenko,
  ``The CNM-hypermultiplet nexus,''
  Nucl.\ Phys.\ B {\bf 543} (1999) 122 [hep-th/9810137].
%
\bibitem{GK2}
  S.~J.~Gates Jr. and S.~M.~Kuzenko,
  ``4D N = 2 supersymmetric off-shell sigma models on the cotangent  
  bundles of  K\"ahler manifolds,''
  Fortsch.\ Phys.\  {\bf 48} (2000) 115
  [hep-th/9903013].
%
\bibitem{ArNi}
  M.~Arai and M.~Nitta,
  ``Hyper-K\"ahler sigma models on (co)tangent bundles with SO(n) isometry,''
  Nucl.\ Phys.\ B {\bf 745}, 208 (2006) 
  [hep-th/0602277].
%
\bibitem{AKL}
  M.~Arai, S.~M.~Kuzenko and U.~Lindstr\"om,
  ``Hyperk\"ahler sigma models on cotangent bundles of Hermitian symmetric
  spaces using projective superspace,''
  JHEP {\bf 0702} (2007) 100
  [hep-th/0612174].
%  
\bibitem{AKL2}
  M.~Arai, S.~M.~Kuzenko and U.~Lindstrom,
  ``Polar supermultiplets, Hermitian symmetric spaces and hyperkahler metrics,''
  JHEP {\bf 0712} (2007) 008
  [arXiv:0709.2633 [hep-th]].
%
\bibitem{KuNo}
  S.~M.~Kuzenko and J.~Novak,
  ``Chiral formulation for hyperkahler sigma-models on cotangent bundles of symmetric spaces,''
  JHEP {\bf 0812} (2008) 072
  [arXiv:0811.0218 [hep-th]].
%
\bibitem{Ku-Co}
  S.~M.~Kuzenko,
  ``On superconformal projective hypermultiplets,''
  JHEP {\bf 0712} (2007) 010
  [arXiv:0710.1479 [hep-th]].
%
\bibitem{K}
  S.~M.~Kuzenko,
  ``Projective superspace as a double punctured harmonic superspace,''
  Int.\ J.\ Mod.\ Phys.\ A {\bf 14} (1999) 1737
  [hep-th/9806147].
%  
\bibitem{KL}
  S.~M.~Kuzenko and W.~D.~Linch,
  ``On five-dimensional superspaces,''
  JHEP {\bf 0602}, 038 (2006)
  [hep-th/0507176].
%
\bibitem{DV}
  F.~Delduc and G.~Valent,
  ``Classical and quantum structure of the compact K\"ahlerian sigma models,''
  Nucl.\ Phys.\ B {\bf 253}, 494 (1985).
%
\bibitem{HiNi}
  K.~Higashijima and M.~Nitta,
  ``Supersymmetric nonlinear sigma models as gauge theories,''
  Prog.\ Theor.\ Phys.\  {\bf 103} (2000) 635
  [hep-th/9911139].
%
\bibitem{KV}
  T.~W.~Kephart and M.~T.~Vaughn,
  ``Tensor Methods For The Exceptional Group E6,''
  Annals Phys.\  {\bf 145} (1983) 162.
%
\bibitem{Springer}
  T.~A.~Springer, Proc. Kon. Ned. Akad. Wet. {\bf A65} (1962) 259.
%
\end{thebibliography}
\end{document}